\documentclass{article}
\usepackage[bindingoffset=0cm,left=3cm,top=2cm,bottom=2cm,right=3cm]{geometry}
\usepackage{authblk}
\usepackage{graphicx}
\usepackage{latexsym}
\usepackage{amsmath,amssymb}
\usepackage{verbatim,times,bbm}

\newtheorem{definition}{Definition}

\begin{document}

\title{Super-Additivity and Entanglement Assistance\\ in Quantum Reading}

\author[1]{Cosmo Lupo}
\affil[1]{York Centre for Quantum Technologies (YCQT), University
of York, York YO10 5GH, UK}
\author[1,2]{Stefano Pirandola}
\affil[2]{Computer Science, University of York, York YO10 5GH, UK}

\maketitle

\begin{abstract}
Quantum information theory determines the maximum rates at which information can be transmitted
through physical systems described by quantum mechanics. Here we consider the communication
protocol known as quantum reading. Quantum reading is a protocol for retrieving the information
stored in a digital memory by using a quantum probe, e.g., shining quantum states of light to
read an optical memory. In a variety of situations using a quantum probe enhances the performances
of the reading protocol in terms of fidelity, data density and energy dissipation.
Here we review and characterize the quantum reading capacity of a memory model, defined as the
maximum rate of reliable reading.
We show that, like other quantities in quantum information theory, the quantum reading capacity
is super-additive. Moreover, we determine conditions under which the use of an entangled ancilla
improves the performance of quantum reading.
\end{abstract}

\section{Introduction}

The scope of quantum information theory is to determine how and
how much information can be stored, processed, and transmitted through
physical systems behaving according to the laws of quantum mechanics \cite{NC}.
In particular, one is interested in transmitting classical or quantum information,
possibly in the presence of physical constraints (e.g., limited energy of
bandwidth) or additional resources (e.g., quantum entanglement or feedback communication) \cite{MW}.

\begin{figure}[t]
\centering
\includegraphics[width=0.4\textwidth]{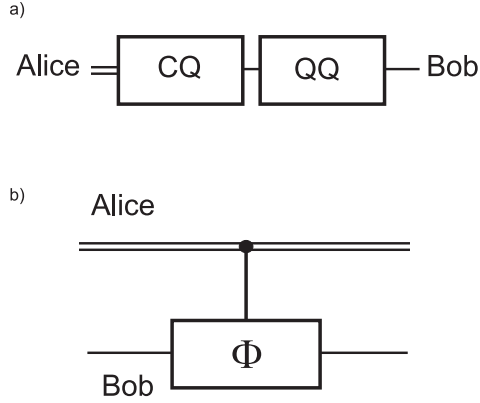}
\caption{a) Classical communication through a QQ channel, where
the CQ channel plays the role of an encoding map. b) In quantum reading
the encoding is represented by a control-QQ channel.}
\label{Fig_schemes}
\end{figure}

In the most common setting, one is given a quantum communication channel,
that is, a physical process that transforms quantum states at the input into
quantum states at the output, as for example an optical fiber does, see e.g., \cite{CD,CGLM}.
A channel of this kind is also called a quantum-quantum (QQ) channel.
Another kind of channel is the so-called classical-quantum (CQ) channel,
which maps classical states (that is, probability distribution over a set
of symbols) into quantum states.
To send classical information through a QQ channel, the sender (Alice) first
encodes classical states into quantum states by applying a suitable CQ channel at the
input of the QQ channel, as represented in Figure \ref{Fig_schemes}a).
In this setting, the CQ channel plays the role of an encoding map.
At the other end of the QQ channel, the receiver (BOB) collects
the output and measures it to decode the classical information sent by Alice.

Quantum reading (QR) is a communication protocol that is based on a different rationale \cite{QR,QRC}.
Instead of having a given communication channel and encoding classical information by
choosing the input states, in QR the sender encodes information by choosing an element
from a collection of QQ channels. Then, to decode, the receiver probes the QQ channels
with a quantum states, collects the output and measures it.

\begin{figure}[t]
\centering
\includegraphics[width=0.4\textwidth]{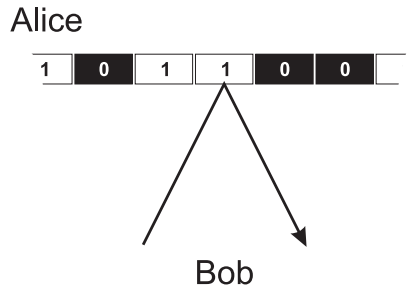}
\caption{An optical memories is the prototype of quantum reading.}
\label{Fig_qread}
\end{figure}

The prototypical example of QR is that of an optical memory, e.g. a CD or DVD, where information is
encoded in a memory cell by means of the physical properties of the substrate, e.g., its reflectivity
or phase.
For this reason, QR has been mainly considered in the context of optical realizations \cite{DellArno,ideal,Tan,Tej,cnoise,Guha,DallArno2014,Roga,Gae}.
For example, a memory cell with low or high reflectivity may encode a logical "0" or "1".
To read this information the receiver must shine a laser beam on the memory cell, and the collect
the reflected beam (see Figure \ref{Fig_qread}).
There are proven advantages in using quantum states of light to perform this task, for instance
increased fidelity and data density, reduced energy consumption and dissipation \cite{QR}.

From a more abstract point of view, QR can be represented as shown in Figure \ref{Fig_schemes}b).
The encoding of a symbol $x$ belonging to an alphabet $\mathcal{X}$ can be modeled as a
control-QQ channel (a generalization of a control-unitary channel \cite{NC})
where the value of $x$ determines which of the QQ channel in a set $\Phi = \{ \phi_x \}_{x \in \mathcal{X}}$
should be applied.

Following \cite{QRC} we refer to the set $\Phi = \{ \phi_x \}_{x \in \mathcal{X}}$ as a ``memory cell''.
One can define the {\it quantum reading capacity} of $\Phi$ as the maximum rate (in bits per use of the memory cell)
that can be reliably transmitted from the sender to the receiver using the encoding procedure specified by $\Phi$.
Indeed, in previous works several notions of capacity have been defined according
to which constraints are assumed or additional resources are allowed \cite{QRC}.
In this paper we present further results concerning quantum reading capacities, in
particular we show that QR capacity is super-additive, and discuss under which conditions
the assistance of entanglement enhances the QR capacity.

The paper proceeds as follows.
In Section \ref{sec:review} we review a few basic notions and definitions.
In Section \ref{sec:nless} we analyze the case of noiseless QR.
The property of super-additivity of QR is discussed in Section \ref{sec:supera},
and the case of noisy QR is considered in Section \ref{sec:noisy}.
Finally, Section \ref{sec:zero} is devoted to zero-error QR capacity, and
Section \ref{sec:end} is for conclusions.

\section{Quantum reading capacities}\label{sec:review}

A QR protocol comprises an encoding and a decoding stage. During
the encoding stage, which is essentially classical, the sender
Alice encodes messages $i=1,2,\dots,M$ using codewords of length
$n$, $x^n(i) = x_1(i)x_2(i) \cdots x_n(i)$, where $x_k(i) \in
\mathcal{X}$. Each codeword identifies a corresponding sequence of
quantum channels from the memory cell $\Phi = \{ \phi_x \}_{x \in
\mathcal{X}}$, e.g., $\phi^n_{x^n(i)} = \phi^n_{x_1(i)} \otimes
\phi^n_{x_2(i)} \otimes \cdots \phi^n_{x_n(i)}$. During the
decoding stage, the receiver Bob prepares a state $\rho^n$, also
called a transmitter, which is used to probe the sequence of
quantum channels $\phi^n_{x^n(i)}$. Finally, Bob collects and
measures the output to retrieve the encoded message.

In analogy with other quantum communication protocol we introduce the following definitions.

\begin{definition}[Quantum Reading protocol]
A $(M, n, \epsilon)$-QR protocol for a memory cell $\Phi = \{ \phi_x \}_{x \in \mathcal{X}}$
is defined by an encoding map $\mathcal{E}$ from $i=1,\dots,M$ to $\mathcal{X}^{\otimes n}$,
a transmitter state $\rho^n$, and a measurement with POVM elements $\{ \Lambda(j)\}_{j \in \mathcal{J}}$,
such that the average probability of error in decoding is less than $\epsilon$, that is,
$$
\frac{1}{M} \sum_i \sum_{j \neq i} \mathrm{Tr} \left[ \Lambda(j) \, \phi^n_{\mathcal{E}(i)}( \rho^n ) \right] \leq \epsilon \, .
$$
The rate of the QR protocol is $R = \frac{1}{n} \log{M}$.
\end{definition}

\begin{definition}[Quantum Reading capacity]
The QR capacity of a given memory cell $\Phi$ is defined as
$$
C(\Phi) = \sup \left\{ \lim_{n}\sup \frac{1}{n} \log{M_n} \, : \, \lim_n \epsilon_n = 0 \right\} \, ,
$$
where the $\sup$ is over all sequences of $(M_n, n, \epsilon_n)$-QR protocols.
\end{definition}

We now introduce some notions of constrained QR capacities.
Instead of considering a generic state for the transmitter,
we can restrict to the family of QR protocols for which the transmitter has the
form, $\rho^n = \rho^{\otimes n}$, that is, it is a separable state across different
uses of the memory cell.
The maximum QR rate that can be achieved under this constraint is defined as $C^1(\Phi)$.
We have
\begin{equation}
C^1(\Phi) = \max_{\rho} C^1(\Phi|\rho) \, ,
\end{equation}
where $C^1(\Phi|\rho)$ denotes the maximum QR rate achievable for a given $\rho$.
Applying known results of quantum information theory \cite{H,SW}, the latter can be
expressed as:
\begin{equation}\label{C1}
C^1(\Phi|\rho) = \max_{ \{ p_x \}_{x \in \mathcal{X}} } \chi \left( \{ p_x, \phi_x(\rho) \}_{x \in \mathcal{X}} \right)
\end{equation}
where the maximum is over all probability distribution over the alphabet $\mathcal{X}$,
\begin{equation}
\chi \left( \{ p_x, \phi_x(\rho) \}_{x \in \mathcal{X}} \right) =
S \left( \sum_{x \in \mathcal{X}} p_x \phi_x(\rho) \right) -
\sum_{x \in \mathcal{X}} p_x \, S \left( \phi_x(\rho) \right) \, ,
\end{equation}
is the Holevo information, and $S \left( \sigma \right) = -
\mathrm{Tr}\left( \sigma \log{\sigma} \right)$ denotes the von
Neumann entropy. It can be easily shown that the maximum is indeed
obtained when $\rho$ is a pure state \cite{QRC}.

Similarly we can define the QR capacities $C^k(\Phi)$, for $k = 1, 2, \dots$,
where the transmitter state is separable across pairs, triplets, etc., of different
uses of the memory cell, that is, $\rho^n = {\rho^k}^{\otimes (n/k)}$ (for $n$ multiple of $k$).
We have
\begin{equation}\label{Ck}
C^k(\Phi|\rho^k) = \frac{1}{k} \max_{ \{ p_{x^k} \}_{x^k \in \mathcal{X}^{\times k}} } \chi \left( \{ p_{x^k}, \phi_{x^k}(\rho^k) \}_{x^k \in \mathcal{X}^{\times k}} \right)
\end{equation}
Clearly $C(\Phi) \geq C^k(\Phi) \geq C^h(\Phi)$ for $k > h$.
If the inequality is strict, that is, $C^k(\Phi) > C^1(\Phi)$ for some $k$,
we say that the $C(\Phi)$ is super-additive.

\begin{figure}[t]
\centering
\includegraphics[width=0.3\textwidth]{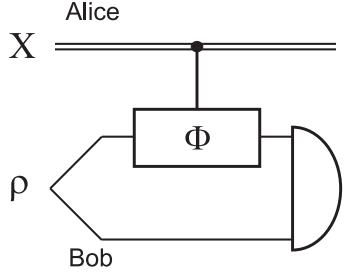}
\caption{Diagram for entanglement-assisted quantum reading.}
\label{QR_EA}
\end{figure}

More generally, the transmitter state can be chosen to be entangled with an
ancilla, which Bob retains and measures jointly, see Figure \ref{QR_EA}.
In this case we speak of entanglement-assisted QR.
Notice that an entanglement-assisted QR protocol for a memory cell $\Phi = \{ \phi_x \}$
is equivalent to an unassisted protocol for the extended memory cell $\Phi \otimes \mathrm{id} = \{ \phi_x \otimes \mathrm{id} \}$,
where $\mathrm{id}$ denotes the identity channel acting on the ancilla.
The entanglement-assisted QR capacity is hence given by the expression
\begin{equation}
C_{EA}(\Phi) = C(\Phi \otimes \mathrm{id}) \, .
\end{equation}
Similarly, we can define the assisted QR capacities $C^k_{EA}(\Phi)$ by
constraining the transmitter to be separable across groups of $k$ uses of the
memory cell.
Clearly, we have $C_{EA}(\Phi) \geq C(\Phi)$, and $C^k_{EA}(\Phi) \geq C^k(\Phi)$.
If, for some value of $k$, this inequality is strict we say that the assistance of entanglement
enhances the QR capacity $C^k(\Phi)$ of the memory cell.

\section{Noiseless quantum reading}\label{sec:nless}

We first consider a noiseless setting in which the QQ channels in the memory cell are unitary
transformations, that is, $\phi_x(\rho) = U_x \rho U_x^\dag$.
For the sake of simplicity we consider the binary setting, $x \in \{0,1\}$, with
the unitaries acting in a finite-dimensional Hilbert space of dimension $d$.

Let us consider the Holevo information,
\begin{align}
\chi\left( \left\{ p_x , U_x |\psi\rangle\langle\psi| U_x^\dag \right\}_{x=0,1} \right) & = S\left[ p \, U_0 \, |\psi\rangle\langle\psi| \, U_0^\dag + (1-p) \, U_1 \, |\psi\rangle\langle\psi| \, U_1^\dag \right] \\
& = S\left[ p \, |\psi\rangle\langle\psi| + (1-p) \, U \, |\psi\rangle\langle\psi| \, U^\dag \right] \, ,
\end{align}
where $p = p_0$, and $U = U_0^\dag U_1$.
From Equation (\ref{C1}) we obtain $C^1$ by maximization of the Holevo information.
This is equivalent to maximizing the von Neuman entropy of the state
$\sigma = p |\psi\rangle\langle\psi| + (1-p) U |\psi\rangle\langle\psi| U^\dag$.
In order to do that, it is convenient to introduce an unit vector $|\psi_\perp\rangle$ such that
$\langle \psi | \psi_\perp \rangle = 0$ and
\begin{equation}
U|\psi\rangle = \alpha |\psi\rangle + \sqrt{1-|\alpha|^2} |\psi_\perp\rangle \, .
\end{equation}
In the system of orthonormal vectors $|\psi\rangle$, $|\psi_\perp\rangle$,
the state $\sigma$ is represented by the density matrix:
\begin{eqnarray}
\tilde\sigma = \left( \begin{array}{cc}
p+(1-p)|\alpha|^2           & (1-p)\alpha\sqrt{1-|\alpha|^2} \\
(1-p)\alpha^*\sqrt{1-|\alpha|^2} & (1-p)(1-|\alpha|^2)
\end{array}\right) \, .
\end{eqnarray}
The maximum von Neumann entropy of $\sigma$ is achieved in correspondence
to the maximum determinant of the matrix $\tilde\sigma$. We have,
\begin{equation}
\det{\tilde\sigma} = p(1-p)(1-|\alpha|^2) \, ,
\end{equation}
which is maximized for $p=1/2$ and in correspondence of the minimum value of $|\alpha|^2 = |\langle \psi | U | \psi \rangle|^2$.
Let us denote as $\{ |j\rangle \}_{j=0,\dots,d-1}$ the eigenvectors of $U$, and as $e^{i\theta_j}$
the corresponding eigenvalues. We expand $|\psi\rangle$ in the basis of eigenvectors,
$|\psi\rangle = \sum_j \psi_j |j\rangle$, which yields $\alpha = \sum_j |\psi_j|^2 e^{i\theta_j}$.
The reading capacity is hence obtained by putting $\alpha = \alpha_{\min}$, with
\begin{equation}\label{amin}
|\alpha_{\min}|^2 = \min_{\{ \psi_j \} | \sum_j |\psi_j|^2=1} \sum_{jj'} \, |\psi_j|^2 \, |\psi_{j'}|^2 \, e^{i(\theta_j-\theta_{j'})} \, ,
\end{equation}
which finally yields
\begin{equation}\label{C1-U}
C^1 = h \left( \frac{1-|\alpha|_{\min}}{2} \right) \, ,
\end{equation}
where $h(x) = -x \log{x} -(1-x) \log{(1-x)}$.

Let us now consider the entanglement-assisted QR capacity $C^1_{EA}$.
To compute $C^1_{EA}$ we can repeat the reasoning of above with $U$ replaced by $U \otimes \mathbb{I}$.
Notice that $U \otimes \mathbb{I}$ has the same eigenvalues of $U$ (but with higher multiplicity).
We can then consider a system of eigenvectors of $U \otimes \mathbb{I}$, denoted as $\{ |jk\rangle \}$,
where $|jk\rangle$ are the eigenvectors with shared eigenvalue $e^{i\theta_j}$.
Expanding the transmitter state $|\psi\rangle$ in this basis we obtain $|\psi\rangle = \sum_{jk} \psi_{jk} |jk\rangle$,
which yields
$\alpha
= \sum_{jk} |\psi_{jk}|^2 e^{i\theta_j}
= \sum_{j} (\sum_k |\psi_{jk}|^2) e^{i\theta_j}$.
We then obtain the same expression for $|\alpha_{\min}|^2$ as in Equation (\ref{amin})
upon replacing $\sum_k |\psi_{jk}|^2 \to |\psi_j|^2$.
In conclusion, we have obtained that $C^1 = C^1_E$, that is, the assistance of
entanglement does not enhance the QR capacity $C^1$ in the noiseless setting.

As an example, let us consider the case of qubit unitaries ($d=2$).
We have $|\psi\rangle = \psi_0 |0\rangle + \psi_1 |1\rangle$ and
\begin{equation}
|\alpha_{\min}|^2 =  \min_{\{\psi_0,\psi_1\} : |\psi_0|^2 + |\psi_1|^2 = 1 }
|\psi_0|^4 + |\psi_1|^4 + 2 |\psi_0|^2 |\psi_1|^2 \cos{(\delta\theta)} \, ,
\end{equation}
with $\delta\theta = |\theta_1-\theta_0|$.
The minimum is hence obtained for $|\psi_0|^2=|\psi_1|^2=1/2$ and yields
\begin{equation}\label{a2}
| \alpha_{\min} | = \sqrt{\frac{1 + \cos{(\delta\theta)} }{2} }
= \left| \cos{\left(\delta\theta/2\right)}\right| \, .
\end{equation}
Finally from (\ref{C1-U}) we obtain
\begin{equation}\label{C1_q}
C^1 = h \left( \sin{\left( \delta\theta/4 \right)}^2 \right) \, .
\end{equation}

\section{Super-additivity}\label{sec:supera}

Let us consider the case of a binary memory cell composed of two qubit unitary transformations, $U_0$ and $U_1$.
We now show that this cell exhibits the phenomenon of super-additivity.

For given $k>1$, let us consider the Holevo information $\chi^k = \chi\left( \left\{ p_{x^k} , \phi_{x^k}(|\psi\rangle\langle\psi|) \right\} \right)$
in Equation (\ref{Ck}) with a probability distribution such that
$p_{00 \cdots 0} = p_{11 \cdots 1} = 1/2$.
That is, we are only considering, with equal probability, the unitary transformations $U_0^{\otimes k}$ and $U_1^{\otimes k}$.
The Holevo information then reads
\begin{equation}
\chi^k = S\left( \frac{1}{2} \, |\psi\rangle\langle\psi| + \frac{1}{2} \, U^{\otimes{k}} |\psi\rangle\langle\psi| {U^{\otimes{k}}}^\dag \right) \, ,
\end{equation}
with $U = U_0^\dag U_1$.

\begin{figure}[t]
\centering
\includegraphics[width=0.5\textwidth]{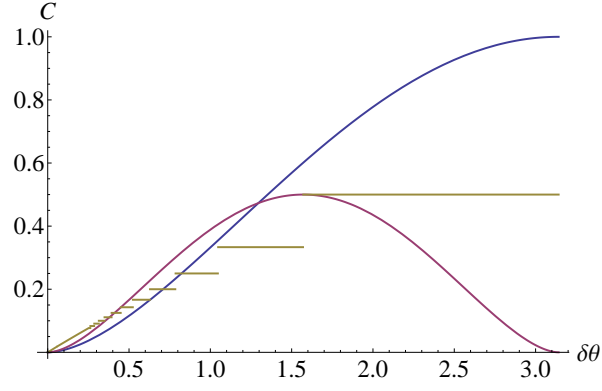}
\caption{QR capacities for a binary qubit unitary memory cell versus the angular separation $\delta\theta$.
Blue line: $C^1$ from Equation (\ref{C1_q}).
Red line: lower bound on $C^2$ from Equation (\ref{Ck_q}).
Green line: zero-error capacity $C^0$ from Equation (\ref{Czero})}
\label{QRC_plot}
\end{figure}

Let us denote the eigenvectors of $U$ as $|0\rangle$ and $|1\rangle$, with corresponding
eigenvalues $e^{i\theta_0}$ and $e^{i\theta_1}$.
As a transmitter state we chose the entangled state $|\psi\rangle = (|0\rangle^{\otimes k} + |1\rangle^{\otimes k})/\sqrt{2}$.
We then obtain
\begin{equation}\label{Ck_q}
C^k \geq \frac{1}{k} \, h \left( \sin{\left( k\delta\theta/4 \right)}^2 \right) \, .
\end{equation}
We hence have found that for any $k$ there exist values of
$\delta\theta$ such that $C^k > C^1$, that is, QR is
super-additive. We can also write a lower bound on the
ultimate QR capacity, that is,
\begin{equation}
C \geq \sup_k
\frac{1}{k} \, h \left( \sin{\left( k\delta\theta/4 \right)}^2 \right) \, .
\end{equation}
See Figure \ref{QRC_plot} for a comparison among different QR
capacities.

\section{Noisy quantum reading}\label{sec:noisy}

Going beyond the case of noiseless quantum reading,
we consider a simple yet physically motivated example of noisy binary quantum reading where
the two encoding maps are of the form
\begin{equation}
\phi_x(\rho) = (1-q) \, U_x \, \rho \, U_x^\dag + q \, \rho_0 \, ,
\end{equation}
where $\rho_0$ is the maximally mixed state in $d$ dimensions, with $x = 0, 1$ and $q \in [0,1]$.

Let us first consider the QR capacity $C^1$. Putting $U = U_0^\dag U_1$, the Holevo information in Equation (\ref{C1}) reads
\begin{align}
\chi & = S\left( (1-q) \left( p_0 |\psi\rangle\langle\psi| + p_1 U |\psi\rangle\langle\psi| U^\dag \right) + q \, \rho_0 \right)
- S\left( (1-q)U|\psi\rangle\langle\psi|U^\dag + q \, \rho_0 \right) \\
& = S\left( (1-q)\left( p_0 |\psi\rangle\langle\psi| + p_1 U |\psi\rangle\langle\psi| U^\dag \right) + q \, \rho_0 \right)
- \eta(1-q+q/d) - (d-1)\eta(q/d) \, , \label{S_1}
\end{align}
where $\eta(y) = -y \log{y}$.
The maximization of the Holevo information is thus reduced to the maximization
of the von Neumann entropy of the state
$\sigma = (1-q)\left( p_0 |\psi\rangle\langle\psi| + p_1 U |\psi\rangle\langle\psi| U^\dag \right) + q \, \rho_0$.
It is convenient to expand this state in a basis defined by the vector $|\psi\rangle$, the vector $|\psi_\perp\rangle$
(such that $\langle \psi | \psi_\perp \rangle = 0$ and $U|\psi\rangle = \alpha |\psi\rangle + \sqrt{1-|\alpha|^2} |\psi_\perp\rangle$),
and any other set of $d-2$ vectors. In this basis the state $\sigma$ is represented by the density matrix
\begin{eqnarray}
\tilde\sigma = \left( \begin{array}{cc}
(1-q)[p+(1-p)|\alpha|^2]+\frac{q}{d}           & (1-q)(1-p)\alpha\sqrt{1-|\alpha|^2} \\
(1-q)(1-p)\alpha^*\sqrt{1-|\alpha|^2} & (1-q)(1-p)(1-|\alpha|^2)+\frac{q}{d}
\end{array}\right) \bigoplus_{j=2}^{d-1} \left( \frac{q}{d} \right) \, .
\end{eqnarray}

The maximum von Neumann entropy of $\sigma$ corresponds to the
maximum determinant of $\tilde\sigma$, where
\begin{equation}
\det{\tilde\sigma} = \left( p (1-p) (1-q)^2 (1-|\alpha|^2) + \frac{q(1-q)}{d} + \frac{q^2}{d^2} \right) \left( \frac{q}{d} \right)^{d-2} \, .
\end{equation}
For any given $q$, the determinant is maximized for $p=(1-p)=1/2$ and in correspondence of the minimum value of $|\alpha|^2$.
We hence obtain
\begin{equation}
C^1 = \eta\left( \frac{q}{d} + \frac{(1-q)(1+|\alpha_{\min}|)}{2}\right)
+ \eta\left( \frac{q}{d} + \frac{(1-q)(1-|\alpha_{\min}|)}{2}\right)
- \eta(1-q+q/d) - \eta(q/d) \, ,
\end{equation}
where $|\alpha_{\min}|$ is given by Equation (\ref{amin}).

For example, in the case $d=2$, using Equation (\ref{a2}), we obtain
\begin{equation}\label{C1-qubit}
C^1 = \eta\left( \frac{q}{2} + (1-q)\cos{(\delta\theta/4)}^2 \right)
+ \eta\left( \frac{q}{2} + (1-q)\sin{(\delta\theta/4)}^2 \right)
- \eta(1-q/2) - \eta(q/2) \, ,
\end{equation}

\subsection{Entanglement assisted reading}

Unlike the noiseless case, the assistance of entanglement can be beneficial in the noisy setting.
We consider the entanglement-assisted QR capacity $C^1_{EA}$, which can be computed by maximization
of the Holevo information
\begin{align}
\chi & = S\left( (1-q)\left( p_0 |\psi\rangle\langle\psi| + p_1 (U \otimes \mathbb{I}) |\psi\rangle\langle\psi| (U^\dag \otimes \mathbb{I}) \right) + q \, \rho_0 \otimes \rho_A \right) \\
& - S\left( (1-q)(U\otimes\mathbb{I})|\psi\rangle\langle\psi|(U^\dag\otimes\mathbb{I}) + q \, \rho_0 \otimes \rho_A \right) \, .
\end{align}
where $|\psi\rangle$ is a joint pure state for the $BA$ system, comprising both Bob output and the ancillary system,
and $\rho_A = \mathrm{Tr}_B(|\psi\rangle\langle\psi|)$ denotes the reduced state of the ancilla.
Without loss of generality we can assume that the dimension of the ancilla $A$
equals that of Bob system $B$.
Moreover, as an example we take $|\psi\rangle$ to be a maximally entangled state in the $BA$ system, which
implies $\rho_A = \rho_0 = \mathbb{I}/d$.
We then obtain
\begin{align}
\chi = & S\left( (1-q) \left( p_0 |\psi\rangle\langle\psi| + p_1 (U \otimes \mathbb{I}) |\psi\rangle\langle\psi| (U^\dag \otimes \mathbb{I}) \right) + q \, \frac{\mathbb{I}}{d^2} \right) \\
& - S\left( (1-q)(U\otimes\mathbb{I})|\psi\rangle\langle\psi|(U^\dag\otimes\mathbb{I}) + q \, \frac{\mathbb{I}}{d^2} \right) \\
= & S\left( (1-q)\left( p_0 |\psi\rangle\langle\psi| + p_1 (U \otimes \mathbb{I}) |\psi\rangle\langle\psi| (U^\dag \otimes \mathbb{I}) \right) + q \, \frac{\mathbb{I}}{d^2} \right) \\
& - \eta(1-q-q/d^2) - (d^2-1)\eta(q/d^2) \, .
\end{align}
We notice that this expression of $\chi$ is formally identical to that in Equation (\ref{S_1})
upon the substitution $d \to d^2$.
We hence obtain
\begin{equation}\label{C1EA_n}
C^1_{EA} \geq
\eta\left( \frac{q}{d^2} + \frac{(1-q)(1+|\alpha|)}{2}\right)
+ \eta\left( \frac{q}{d^2} + \frac{(1-q)(1-|\alpha|)}{2}\right)
- \eta(1-q+q/d^2) - \eta(q/d^2) \, ,
\end{equation}
where $|\alpha|^2 = |\langle \psi | U \otimes \mathbb{I} | \psi \rangle|^2$ and
$|\psi\rangle$ is a maximally entangled state.

For the sake of simplicity, let us now consider the case $d=2$
and consider a system of eigenvectors of $U \otimes \mathbb{I}$,
denoted as $\{ |jk\rangle \}$, where $|jk\rangle$ is an eigenvector with eigenvalue $e^{i\theta_j}$.
The maximally entangled state $\psi$ can be represented, without loss of generality, as
$|\psi\rangle = \sum_{j=0,1} 2^{-1/2} |jj\rangle$, which implies
$\alpha = \frac{1}{2} \sum_{j=0,1} e^{i\theta_j}$ and in turn yields
$|\alpha| = |\cos{(\delta\theta/2)}|$.
Substituting this value for $\alpha$ in Equation (\ref{C1EA_n}) with $d=2$ we obtain
\begin{equation}\label{C1EA-qubit}
C^1_{EA} \geq \eta\left( \frac{q}{4} + (1-q) \cos{(\delta\theta/4)}^2 \right)
+ \eta\left( \frac{q}{4} + (1-q) \sin{(\delta\theta/4)}^2 \right)
- \eta(1-3q/4) - \eta(q/4) \, .
\end{equation}

\begin{figure}[t]
\centering
\includegraphics[width=0.7\textwidth]{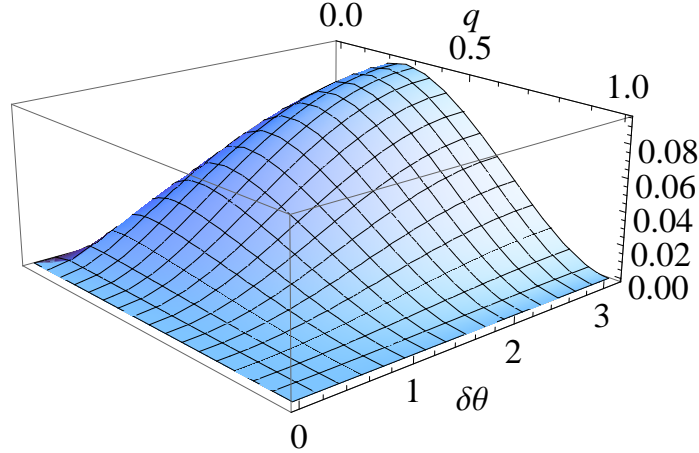}
\caption{The plot shows a lower bound on the gain $C^1_{EA} - C^1$
for the case of the qubit binary noisy memory cell described in Section \ref{sec:noisy}.
The lower bound is obtained by taking the difference between the expressions in Equation (\ref{C1EA-qubit})
and Equation (\ref{C1-qubit}).}
\label{dC}
\end{figure}

By comparison with Equation (\ref{C1-qubit}) it follows that, unlike the noiseless setting, the assistance of entanglement is
beneficial (that is, $C^1_{EA} > C^1$) in the presence of noise (see Figure \ref{dC}).

\section{Zero-error capacity}\label{sec:zero}

QR is closely related to the problem of quantum channel discrimination \cite{QR,QRC}.
The particular case of noiseless QR is hence in close relation with the problem of discriminating
between two unitary transformations.
According to \cite{Acin}, two unitaries , $U_0$ and $U_1$,
can always be perfectly discriminated, if enough copies of them are provided and using a
suitable input state and, possibly, a collective measurement.
To relate in a formal way this feature of the problem of unitary discrimination with
QR, we need to consider the notion of zero-error QR capacity.

\begin{definition}[Zero-error quantum reading protocol]
A $(M, n)$ zero-error QR protocol for a memory cell $\Phi = \{ \phi_x \}_{x \in \mathcal{X}}$
is defined by an encoding map $\mathcal{E}$ from $i=1,\dots,M$ to $\mathcal{X}^{\otimes n}$,
a transmitter state $\rho^n$, and a measurement with POVM elements $\{ \Lambda(j)\}_{j \in \mathcal{J}}$,
such that the average probability of error in decoding is zero.
The rate of the QR protocol is $R = \frac{1}{n} \log{M}$.
\end{definition}

From the definition of zero-error QR protocol it follows that of zero-error QR capacity as
the maximum zero-error QR rate.
Coming back to the results of \cite{Acin}, let us consider the case of two-dimensional unitaries
(extension to the higher dimension is straightforward).
Given that the spectrum of the unitary $U = U_0^\dag U_1$ is $e^{i \theta_0}$, $e^{i \theta_1}$,
\cite{Acin} proved that the unitaries $U_0^{\otimes n}$ and $U_1^{\otimes n}$ are perfectly
distinguishable if $n \geq \pi/\delta\theta$ (with $\delta\theta = |\theta_0 - \theta_1|$).
This result implies that the zero-error QR capacity of the binary unitary memory cell is
\begin{equation}\label{Czero}
C_0 = \frac{1}{\lceil\pi/\delta\theta\rceil} \, .
\end{equation}
The zero-error capacity is plotted in Figure \ref{QRC_plot}.

\section{Conclusions}\label{sec:end}

We have presented several new results concerning the properties of
super-additivity and the use of entanglement as a resource to
enhance the QR capacities. We have proven that the assistance of
entanglement does not increase the QR capacity in the noiseless
setting, where the memory cell consists of unitary
transformations. On the contrary, we have shown with an example
that the assistance of entanglement may enhance the QR capacity
when the memory cell consists of noisy quantum channels. We have
also shown that the QR capacity, like other quantities in quantum
information theory \cite{Hastings,supera}, exhibits the phenomenon
of super-additivity.

As already pointed out in previous works \cite{QR, QRC}, the protocol of QR is closely related
to the task of quantum channel discrimination.
For example, the fact that the assistance of entanglement enhances the QR capacity for a noisy memory cell,
mirrors the fact that the use of an entangled ancilla may improve the discrimination between quantum channels (see e.g. \cite{Sacchi}).
At a more formal level, the analogy between QR and quantum channel discrimination can be
appreciated through the notion of zero-error QR capacity, as discussed in Section \ref{sec:zero}.
QR is also closely related to the task of parameter estimation, see e.g. \cite{Paris} 
(this connection was also discussed in \cite{rafal}).
For example, notice that the QR capacity is super-additive in the region of small values of
$\delta\theta$ (see Figure \ref{QRC_plot}). This is the regime in which discriminating between
two unitaries is essentially equivalent to estimating a small variation of the value of a relative phase.

\section*{Acknowledgements}
S.P. has been supported by the EPSRC (`qDATA', EP/L011298/1).

\end{document}